\documentclass[reprint,superscriptaddress,nofootinbib,amsmath,amssymb,aps,floatfix]{revtex4-1}
\pdfoutput=1
\usepackage{graphicx}
\usepackage{dcolumn}
\usepackage{bm}
\usepackage[utf8]{inputenc}
\usepackage{graphics,graphicx}
\usepackage{parskip}
\usepackage{amsmath,amssymb}
\usepackage[normalem]{ulem}
\usepackage{amsfonts}

\usepackage{multirow}
\usepackage{fancyhdr}
\usepackage{xcolor}
\usepackage{placeins}
\usepackage[title]{appendix}
\usepackage{wasysym}
\usepackage{url}
\usepackage{braket}
\usepackage{siunitx}
\usepackage{soulpos}
\usepackage{float}
\usepackage[caption=false]{subfig}
\usepackage{lipsum}
\usepackage{booktabs}
\usepackage{comment}
\usepackage{braket}
\usepackage[colorlinks=true,linkcolor=red,citecolor=blue,filecolor=.]{hyperref}
\usepackage{cleveref}
\usepackage{natbib}
\usepackage{setspace}
\usepackage{subfiles} 
\usepackage{upgreek}

\begin{document}
\newcommand{\matt}[1]{\textcolor{green}{#1}}
\newcommand{\ematt}[1]{\textcolor{blue}{#1}}

\newcommand{\dan}[1]{\textcolor{magenta}{#1}}
\newcommand{\dandan}[1]{\textcolor{orange}{#1}}
\newcommand{\epeter}[1]{\textcolor{blue}{#1}}
\title{Metastable doubly-charged Rydberg molecules}

\author{Daniel J. Bosworth} 
 \email{dboswort@physnet.uni-hamburg.de}
\affiliation{%
 Zentrum f\"ur Optische Quantentechnologien, Universit\"at Hamburg, Luruper Chaussee 149, 22761 Hamburg, Germany\\
}%
\affiliation{%
 The Hamburg Centre for Ultrafast Imaging, Universit\"at Hamburg, Luruper Chaussee 149, 22761 Hamburg, Germany\\
}%
\author{Matthew T. Eiles}%
\affiliation{%
 Max-Planck-Institut für Physik komplexer Systeme, Nöthnitzer Straße 38, 01187 Dresden, Germany\\
}%
\author{Peter Schmelcher}
\affiliation{%
 Zentrum f\"ur Optische Quantentechnologien, Universit\"at Hamburg, Luruper Chaussee 149, 22761 Hamburg, Germany\\
}%
\affiliation{%
 The Hamburg Centre for Ultrafast Imaging, Universit\"at Hamburg, Luruper Chaussee 149, 22761 Hamburg, Germany\\
}%

\date{\today}

\begin{abstract}
H$_3^{2+}$ is a one-electron system with three positive nuclei and is known to be unstable in its electronic ground-state. We examine an analogous one-electron system composed of a $^{87}$Rb Rydberg atom interacting with a pair of cations and predict the existence of metastable vibrationally-bound states of $^{87}$Rb$_3^{2+}$. These molecules are long-range trimers whose stability rests on the presence of core-shell electrons and favourable scaling of the Rydberg atom's quadrupole moment with the principal quantum number $n$. Unlike recently observed ion-Rydberg dimers, whose binding is due to internal flipping of the Rydberg atom's dipole moment, the binding of $^{87}$Rb$_3^{2+}$ arises from the interaction of the ions with the Rydberg atom's quadrupole moment. The stability of these trimers is highly sensitive to $n$. We do not expect these states to exist below $n=24$ and for $n \leq 35$, their lifetime is limited by tunnelling of the Rydberg electron. In contrast, at very large $n$ the lifetime will be limited by tunnelling of the vibrational wavepacket. In between these limits, we expect a range of bound states at intermediate $n$ for which both tunnelling rates are smaller than the radiative decay rate of the Rydberg state.
\end{abstract}

\maketitle
\section{Introduction}~\label{sec:intro}
Atomic and molecular ions play significant roles in chemical processes throughout nature. One example can be found inside dense interstellar gases, where proton transfer reactions between H$_3^{+}$ molecular ions and neutral species contribute to the synthesis of, among other things, water~\cite{Herbst2000Astrochemistry,Petrie2007Ions}. Despite their importance, molecular ions are generally short-lived. This is especially true for multiply-charged variants, whose decay releases considerable amounts of stored molecular energy through a Coulomb explosion between their singly-charged fragments~\cite{Falcinelli2020Production}. This highly-exothermic decay has in the past inspired proposals to use molecular dications as a source of propulsive energy~\cite{Nicolaides1989Energy}.\par
Molecular ions may also form within the ultracold environment of a Bose-Einstein condensate~\cite{Cote2002Mesoscopic,Schurer2014Ground,Schurer2017Unravelling}. Three-body processes between neutral atoms and an atomic ion can lead to spontaneous formation of bound atom-ion dimers which occupy rovibrational states close to the dissociation threshold. By virtue of the long-range nature of the atom-ion interaction~\cite{Cote2000From}, such states are mesoscopic in size, possessing bond lengths on the order of 0.1 $\upmu$m. These molecules have been well-characterised theoretically~\cite{Cote2002Mesoscopic,Gao2010Universal,Idziaszek2011Quantum,Schurer2014Ground,Schurer2017Unravelling,Bosworth2021Spectral} and weakly-bound states of $^{88}$Sr$^+$-$^{87}$Rb were recently observed in a hybrid trap setup~\cite{Pinkas2023Trap}.\par
By promoting atoms to highly-excited states, it is even possible to form \textit{macroscopic} molecular ions, where a single ion binds a Rydberg atom~\cite{Duspayev2021Long,Deiss2021Long,Zuber2022Observation}. Similar to other species in the zoo of long-range Rydberg molecules~\cite{Greene2000Creation,Bendkowsky2009Observation,Fey2019Ultralong,Sassmannshausen2016Long,Hollerith2023Rydberg,Schmidt2016Mesoscopic,Schmidt2018Theory,Rittenhouse2010Ultracold,GonzalezFerez2015Rotational}, Rydberg molecular ions exhibit micrometre-size bond lengths and binding energies on scales of mega-/gigahertz for sufficiently large $n$. The macroscopic size and slow vibrational dynamics of ion-Rydberg systems allow the imaging of molecular dynamics without the need for ultrafast pulses~\cite{Zou2023Observation,Berngruber2024Insitu}. More generally, the high density of states in Rydberg molecules mean they can be utilised to explore and exploit effects beyond the Born-Oppenheimer approximation~\cite{Hummel2021Synthetic,Hummel2023Vibronic,Srikumar2023Nonadiabatic,Eiles2024Katos}.\par
In this paper, we extend prior work on interacting ion-Rydberg pairs by exploring a three-body system of a $^{87}$Rb Rydberg atom interacting with two $^{87}$Rb$^+$ cations. This system bears resemblance to the one-electron H$_3^{2+}$ problem, which is expected to be unstable in its electronic ground-state~\cite{Conroy1964Results,Medel2008Nonexistence,Glasser2014Quantum} and has so far eluded experimental detection~\cite{Berkowitz1971Search}. Despite the strong Coulomb repulsion between the cations, we find that the presence of quantum defect states~\footnote{Quantum defect states guarantee significant avoided crossings between potential energy surfaces in the electronic structure. For our needs however, the precise values of the quantum defect parameters are unimportant.} and the Rydberg atom's large quadrupole moment give rise to potential wells supporting several vibrationally-bound states of $^{87}$Rb$_3^{2+}$ above a critical value of the principal quantum number $n$. Put differently, we reveal that introducing additional energy into the system through a Rydberg excitation can unexpectedly lead to its stabilisation. We further explore possible decay mechanisms and highlight a range of $n$ for which we expect these trimers to be stable over the radiative lifetime of the Rydberg atom. \par
This work is laid out as follows. In section~\ref{sec:scaling}, we discuss ion-Rydberg interactions and derive a lower-bound for $n$ above which we expect the system to support binding potentials. Section~\ref{sec:numerics} then presents numerical results for adiabatic potential energy surfaces of the system as well as vibrationally-bound states. The possible decay mechanisms of $^{87}$Rb$_3^{2+}$ and their variation with $n$ is the focus of section ~\ref{sec:decay}.
\section{Setup \& Interactions}\label{sec:scaling}
In this section, we first describe the $^{87}$Rb$_3^{2+}$ system with its competing ion-ion and ion-Rydberg interactions. We then demonstrate that for sufficiently large $n$ the ion-Rydberg interaction can be strong enough to counterbalance the destabilising Coulomb repulsion of the ion pair.\par
\begin{figure}
    \centering
    \includegraphics[width=0.475\textwidth]{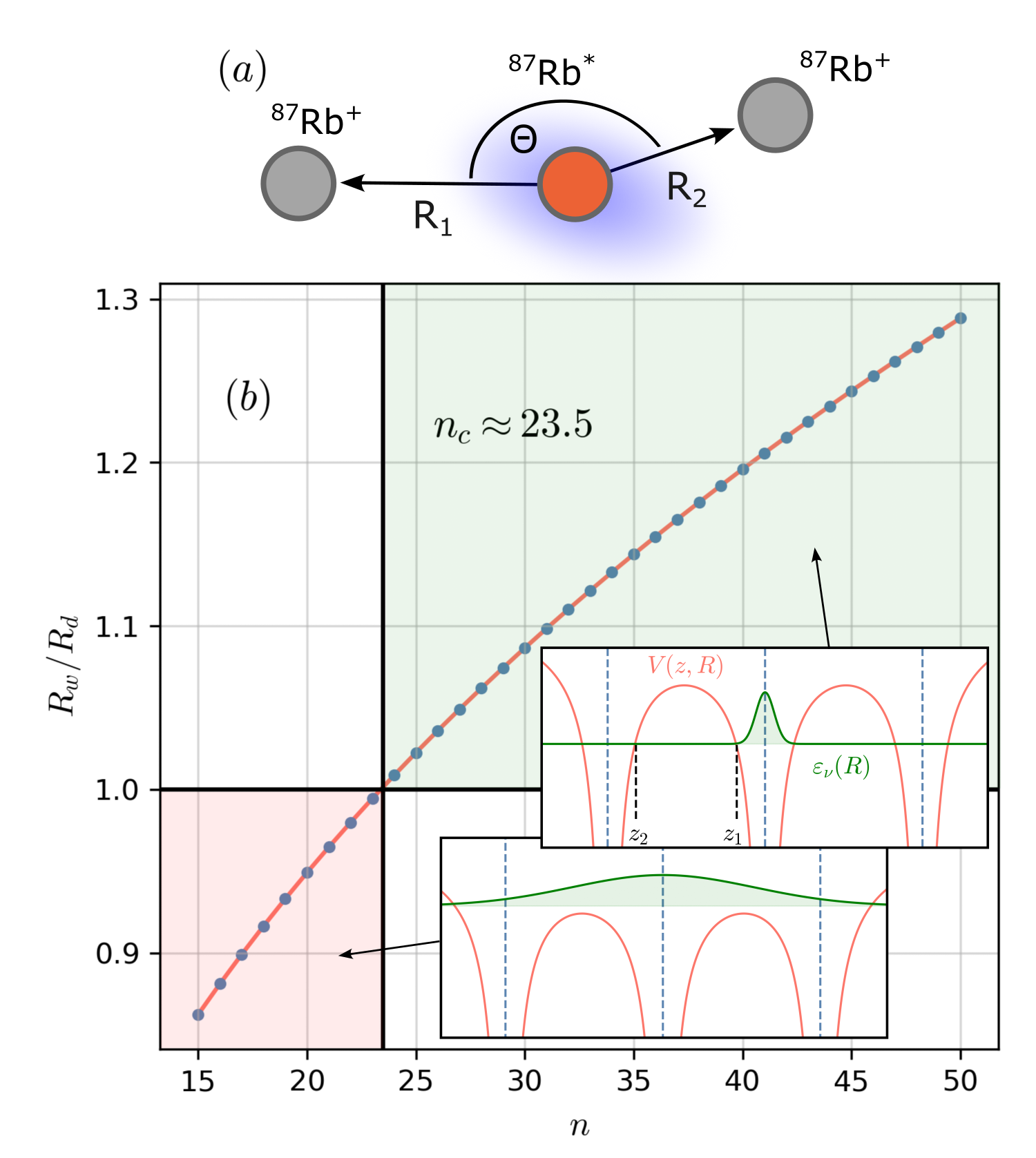}
    \caption{(a) Schematic of the $^{87}\text{Rb}_3^{2+}$ system, consisting of two $^{87}\text{Rb}^{+}$ ions and an $^{87}$Rb Rydberg atom.
    (b) The scaling of the critical separations $R_w$ and $R_d$ with $n$. We find $R_w/R_d=1$ for a critical $n$ value of $n_c\approx 23.5$. The solid line is a fit to a power-law function $a n^b$, with $a = 0.350$ and $b=0.333$ to three significant figures. The insets depict how the energy of the Rydberg electron relates to the maximum height of the Coulomb barrier in the regimes $R_w/R_d < 1$ and $R_w/R_d > 1$.}
    \label{fig:intro-fig}
\end{figure}

The system under consideration is shown in figure~\ref{fig:intro-fig}~(a). We consider a single Rydberg atom interacting with two positively-charged ions at positions $\mathbf{R}_1 = (R_1,0,0)$ and $\mathbf{R}_2 = (R_2,\Theta,0)$ relative to the Rydberg core. The electronic Hamiltonian is given by $\hat{H}_e = \hat{H}_0 + \hat{V}_{\text{IR}}^{(1)}(\mathbf{R}_1) +  \hat{V}_{\text{IR}}^{(2)}(\mathbf{R}_2) + \hat{V}_{\text{II}}(\mathbf{R}_1,\mathbf{R}_2)$. $\hat{H}_0$ describes the Rydberg atom, $\hat{V}_{\text{IR}}^{(i)}$ is the interaction of the Rydberg atom with the $i^{\text{th}}$ ion and $\hat{V}_{\text{II}} = 1/|\mathbf{R}_1-\mathbf{R}_2|$ is the Coulomb repulsion between the ions. Unless stated otherwise, we assume atomic units throughout this work. We further neglect the fine structure of the Rydberg atom in our calculations.\par
For internuclear separations exceeding the Le Roy radius~\cite{LeRoy1974Long} where the overlap between the separate atomic charge distributions vanishes, the ion-Rydberg interaction can be expressed as a multipole expansion~\cite{Jackson1998Classical}:
\begin{equation}\label{eq:ion-ryd-int}
    \hat{V}_{\text{IR}}^{(i)}(\mathbf{R}_i) = -\sum_{\lambda=1}^{\infty}\sum_{\mu=-\lambda}^{+\lambda}\frac{4\pi}{2\lambda+1}\frac{r^\lambda}{R_i^{\lambda+1}}Y_{\lambda\mu}(\theta,\phi)Y_{\lambda\mu}^*(\Theta,0).
\end{equation}
Where $\mathbf{r} = (r,\theta,\phi)$ denotes the position of the Rydberg electron relative to the Rydberg core and $Y_{\lambda\mu}$ are spherical harmonics. Regarding the arrangement of the nuclei, in this work we are particularly interested in the symmetric case in which the ions are located either side of the Rydberg atom with internuclear spacing $|\mathbf{R}_1|=|\mathbf{R}_2|=R$ and $\Theta=\pi$. In this symmetric linear arrangement, the odd-$\lambda$ terms in $\hat{V}_{\text{IR}}^{(1)}$ exactly cancel out those in $\hat{V}_{\text{IR}}^{(2)}$, such that the leading order term in the net ion-Rydberg interaction is the interaction of the ions with the Rydberg atom's quadrupole moment $V_{\text{quad}}(R)\propto - n^4/R^3$. This setup thus offers the possibility to explore physics dominated by quadrupole interactions. \par
Potential wells accommodating bound states may form in purely long-range molecular systems due to avoided crossings between different potential energy curves (PEC). For ion-Rydberg dimers~\cite{Duspayev2021Long,Deiss2021Long}, these avoided crossings occur between a high-field seeking $\ket{nP}$ state and low-field seeking high angular momentum states $\ket{(n-1)l>3}$ whose leading-order energy corrections are due to the interaction with the Rydberg atom's dipole moment (the charge-dipole term $\lambda=1$ in the series~\eqref{eq:ion-ryd-int}). Their anticrossings form binding potentials at internuclear separations approximately one order of magnitude larger than the extent of the Rydberg electron's orbit.
The $^{87}$Rb$_3^{2+}$ system might also exhibit such crossings between high- and low-field seeking states, possibly giving rise to binding potentials. However, in this case the ion-Rydberg interaction must contend with the Coulomb repulsion of the ion pair and may thus be washed out. Nevertheless, given that the strength of $V_{\text{quad}}(R)$ scales with $n$, is there a range of $n$ for which the strength of this interaction is comparable to that of the Coulomb interaction?\par
To answer this question, let us consider two Rydberg states $\psi_A$ and $\psi_B$ connected asymptotically ($R_1,R_2\rightarrow\infty$) to neighbouring hydrogenic manifolds $n$ and $n-1$. In the absence of the ion pair, these states have energies $\varepsilon^{(0)}_A$ and $\varepsilon^{(0)}_B$, such that $\varepsilon^{(0)}_A > \varepsilon^{(0)}_B$. 
In the presence of the ion pair, the states experience energy shifts:
\begin{equation}~\label{eq:energy-correction}
    \Delta\varepsilon^{(1)}_j(R) = \frac{p_j\, n^4}{R^3} + \frac{1}{2R}.
\end{equation}
The first term represents the leading-order energy correction resulting from the interaction of the ions with the quadrupole moment of the Rydberg atom $\braket{\psi_j|V_{\text{quad}}(R)|\psi_j}$. This correction lifts the state degeneracy within the hydrogenic manifold. The second term is the energy correction arising from the Coulomb repulsion between the ions $\hat{V}_{\text{II}}$, which contributes a global energy shift of all states.
If the energy correction $\braket{\psi_A|V_{\text{quad}}(R)|\psi_A}$ is negative with a magnitude larger than $\braket{\psi_B|V_{\text{quad}}(R)|\psi_B}$, $\psi_A$ can in principle become degenerate with $\psi_B$ at a sufficiently small separation $R = R_w$ where $|\braket{\psi_A|V_{\text{quad}}(R_w)|\psi_A}| \approx \varepsilon^{(0)}_A - \varepsilon^{(0)}_B$. Corrections beyond equation~\eqref{eq:energy-correction} ensure however that this degeneracy is prohibited. The resulting level-repulsion between $\psi_A$ and $\psi_B$ may then lead to the formation of a binding potential.\par
Now we consider how the crossing point $R_w$ of the uncoupled energy curves should scale with $n$. Setting $\varepsilon^{(0)}_A + \Delta\varepsilon^{(1)}_A(R_w) = \varepsilon^{(0)}_B + \Delta\varepsilon^{(1)}_B(R_w)$ and knowing that the energy spacing between neighbouring Rydberg states decreases according to $\varepsilon^{(0)}_A - \varepsilon^{(0)}_B = q \,n^{-3}$, with $q$ a positive constant, we find:
\begin{equation}~\label{eq:Rw}
    R_w = \frac{2p}{q}n^{7/3},
\end{equation}
where we assume that $|p_A| = |p_B|$.\par
Since we are interested in long-range bound molecular states, the Rydberg electron should remain localised on the central positive core. For three fixed positive ions centred at the origin with equal spacing $|\mathbf{R}|$, the Rydberg electron experiences the net Coulomb potential $V(\mathbf{r}) = -1/|\mathbf{r}|-1/|\mathbf{r}+\mathbf{R}| -1/|\mathbf{r}-\mathbf{R}| + 5/2|\mathbf{R}|$, where the final term is the total repulsive interaction between the positive ions. The maximum height of the Coulomb barrier separating the electron from the two surrounding ionic cores occurs at $r = R/2$, giving a barrier height of $V_b(R) = -13/6R$. Below a critical internuclear separation $R_d$, the energy of the Rydberg electron will exceed $V_b(R)$ such that it delocalises over all three ions, potentially destabilising the system. Setting $\varepsilon^{(0)}_A + \Delta\varepsilon^{(1)}_A(R_d) = V_b(R_d)$, we find:
\begin{equation}~\label{eq:Rd}
    3R_d^3 - 16n^2 R_d^2 + 6 p n^6 = 0.
\end{equation}
We now compare the scaling of $R_w$ and $R_d$ with $n$. For this, we first determine values for the constants $p$ and $q$. $p$ is determined by fitting the radial integral appearing in $\braket{\psi_j|V_{\text{quad}}(R)|\psi_j}$ to the power-law function $a n^b$~\footnote{We find $a \approx 2.09$ and $b\approx 4.04$. Given $\braket{\psi_j|V_{\text{quad}}(R)|\psi_j} = \sqrt{16\pi/5}\braket{\psi_j|r^2 Y_{2,0}(\theta,\phi)|\psi_j}/R^3 = p n^b/R^3$, this yields $p = 1.06$. Values are given to three significant figures.}, whilst $q$ is found by fitting the energy splitting of neighbouring Rydberg manifolds to $q n^c$~\footnote{We find $q\approx 1.16$ and $c\approx -3.03$ to three significant figures.}.\par
Figure~\ref{fig:intro-fig} (b) shows the ratio of the two critical separations $R_w/R_d$ as a function of $n$. The values of $R_d$ were determined through numerical solution of~\eqref{eq:Rd}. The ratio is foudn to be monotonically increasing over the range of $n$ shown and exceeds unity for $n \geq 24$. This serves as a lower bound above which $^{87}$Rb$_3^{2+}$ may exhibit binding potentials in which the Rydberg electron is localised on the central $^{87}$Rb$^{+}$ ion.\par
The increase of $R_w/R_d$ with $n$ can be understood by considering~\eqref{eq:Rd} in the limit of large $n$, for which it may be approximated as $-16 n^2 R_d^2 + 6p n^6 \approx 0$. This yields $R_d \propto n^2$, from which we obtain $R_w/R_d \propto n^{1/3}$. This result agrees with the exponent of the power-law fit to the data points in figure~\ref{fig:intro-fig} (b) (see caption).
\begin{figure}
    \centering
    \includegraphics[width=0.475\textwidth]{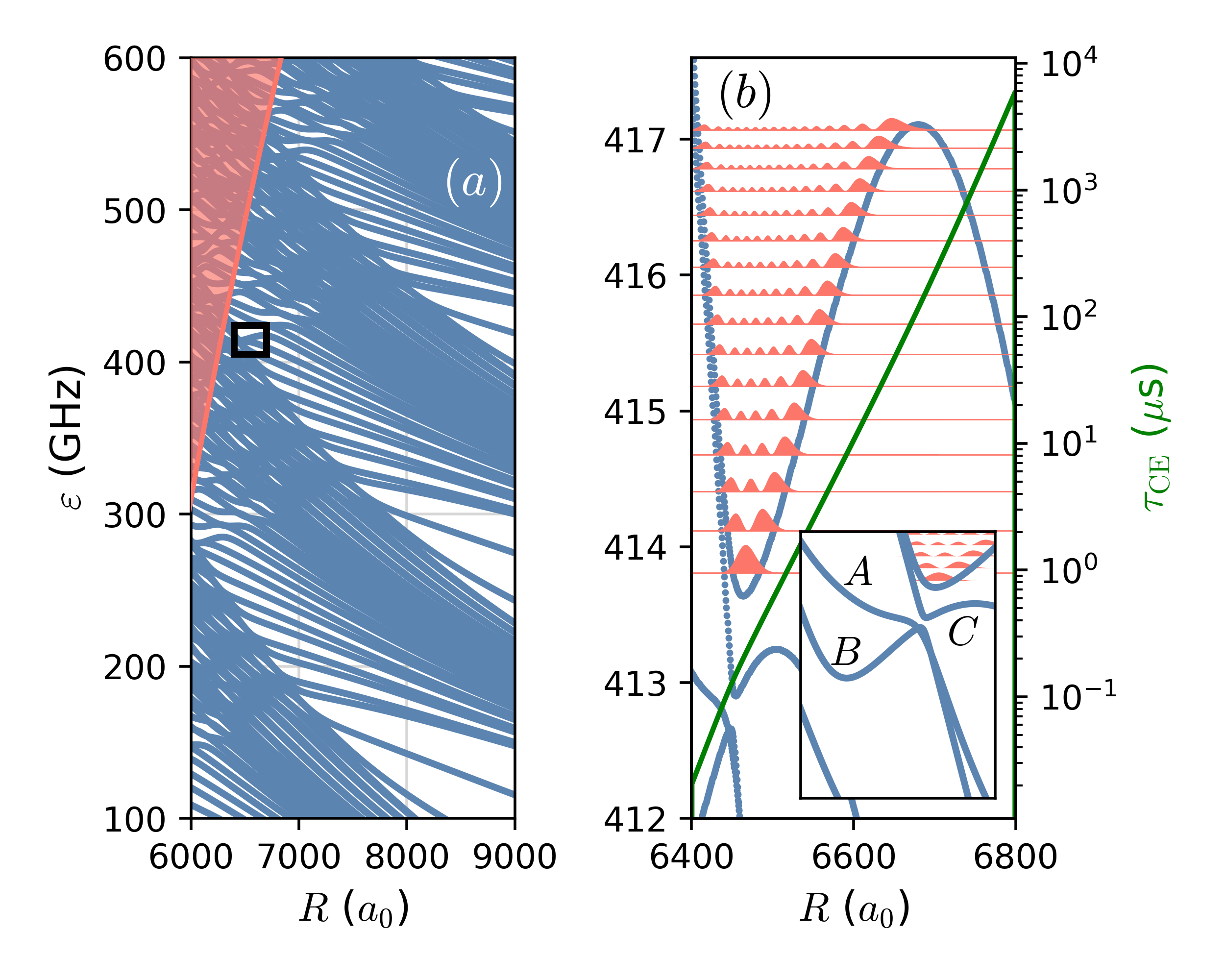}
    \caption{
Adiabatic PES near the $n=35$ hydrogenic manifold. (a) 1D slice along PES for the symmetric linear configuration $R_1 = R_2 = R,\;\Theta=\pi$. The shaded region in the upper left indicates the region in which the localised electron ansatz no longer holds. (b) Magnification of the box in (a) showing an example of a binding potential. The filled curves show the probability density of the symmetric stretching modes determined from the effective 1D vibrational Hamiltonian~\eqref{eq:vib-ham-1D}. The solid green line on the second $y$-axis is the $R$-dependent charge exchange time of the Rydberg electron in $\upmu$s determined using the WKB approach (see section~\ref{sec:decay} and appendix~\ref{sec:wkb} for details). Inset shows the three neighbouring adiabatic PES of the binding potential, labelled $A-C$. Energies given in GHz relative to the 35H atomic Rydberg state.}
    \label{fig:pes}
\end{figure}
\section{Electronic and vibrational structure}\label{sec:numerics}
We now diagonalise the electronic Hamiltonian $\hat{H}_e$ in a finite basis of atomic Rydberg states fulfilling $\hat{H}_0\phi_{\alpha}(\mathbf{r}) = \epsilon_{\alpha}\phi_{\alpha}(\mathbf{r})$. The eigenvalues $\varepsilon_{\nu}(\mathbf{R})$ of $\hat{H}_e$ depend on the system's three internal degrees of freedom $\mathbf{R}= (R_1,R_2,\Theta)$ and form potential energy surfaces (PES) which describe the interaction potential between the nuclei. Both the Rydberg Hamiltonian $\hat{H}_0$ and the Coulomb interaction operator $\hat{V}_{\text{II}}(\mathbf{R}_1,\mathbf{R}_2)$ are diagonal in the Rydberg basis. Matrix elements of the ion-Rydberg interaction terms $\hat{V}_{\text{IR}}^{(i)}(\mathbf{R}_i)$ can be evaluated straightforwardly in the Rydberg basis, since analytical results for integrals of multipole moments are generally available~\cite{Friedrich2017Theoretical}. We include the first sixteen $\lambda$ terms in the multipole expansion~\eqref{eq:ion-ryd-int}.\par
In figure~\ref{fig:pes}, we show one-dimensional cuts through the adiabatic PES near the $n=35$ hydrogenic manifold for symmetric linear configurations of the nuclei where $R_1=R_2=R$ and $\Theta=\pi$. Figure~\ref{fig:pes} (a) shows the adiabatic PES at the onset of the Inglis-Teller regime~\cite{Inglis1939Ionic}, where states from different manifolds are mixed~\cite{Gallagher2003Back}. The shaded region indicates electronic states with $\varepsilon_{\nu}(R) \geq V_b(R)$, such that the Rydberg electron delocalises over all three positive cores. In this regime, our basis of Rydberg states localised on the central positive core is no longer an accurate description of the system. To that end, we restrict our focus to regions for which $\varepsilon_{\nu}(R) < V_b(R)$ holds.\par
At large $R$, the dominant energy correction to the eigenstates is a global positive energy shift stemming from the Coulomb repulsion between the ions. In figure~\ref{fig:pes}~(a), we see that at $R=9000~a_0$ this shift amounts to approximately 400~GHz. This can be seen by looking at the energy of the states in the second manifold from the top of the subplot, which connect asymptotically to the $n=35$ hydrogenic manifold.\par
Despite this large global energy shift, we see that there is also a significant Stark-like splitting of the states in the manifolds due to the interaction of the Rydberg atom with the electric field of the ion pair.
For approximately $R < 7500~a_0$, the splitting becomes comparable to the energy separation of the manifolds and the resultant avoided crossings form a variety of well-like structures. One such well is shown in figure~\ref{fig:pes}~(b). In general, the wells in this region of the energy spectrum have depths ranging from hundreds of MHz to several GHz with local minima positioned at internuclear separations two to three times larger than the expected radius of the Rydberg electron's orbit. 
\begin{figure}[t]
    \centering
    \includegraphics[width=0.475\textwidth]{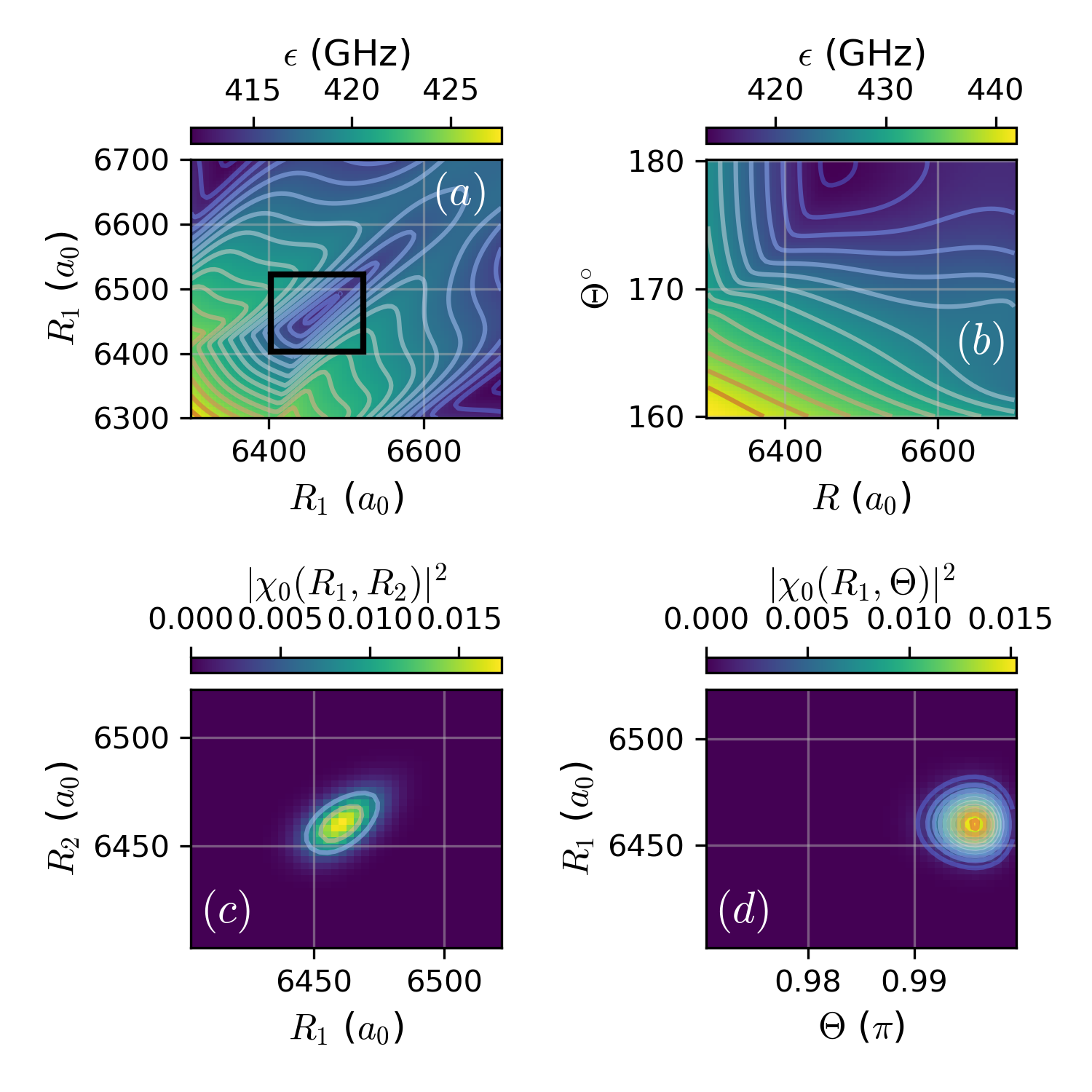}
    \caption{ (a), (b) 2D slices of the binding potential shown in figure~\ref{fig:pes}~(b) for $\Theta=\pi$ and $R_1=R_2=R$. The adiabatic PES exhibits a local minimum for $R_1=R_2=6460~a_0$ and $\Theta=\pi$. Energies given in GHz relative to the $35$H atomic Rydberg state. (c),(d) Reduced probability density of the vibrational ground-state $\chi_0(R_1,R_2,\Theta)$ averaged over (c) $\Theta$ and (d) $R_2$. The states shown here are normalised as $\int d R_1 d R_2 d \Theta |\chi(R_1,R_2,\Theta)|^2=1$. The size of plot (c) is illustrated by the black box shown in (a) for scale comparison.}
    \label{fig:2d-pes}
\end{figure}
In comparison, the binding potentials of diatomic ion-Rydberg molecules arise at internuclear separations roughly one order of magnitude larger than the Rydberg electron's orbit~\cite{Deiss2021Long,Zuber2022Observation}. Consequently, we expect that charge exchange processes will be more important in our system and we will explore this later in section~\ref{sec:decay}.\par
Figure~\ref{fig:2d-pes}~(a) and~(b) show two-dimensional slices of the adiabatic PES of the well in figure~\ref{fig:pes} for non-symmetric nuclear configurations. These plots indicate that the well in figure~\ref{fig:pes}~(b) is indeed a local minimum for all internal degrees of freedom of the triatomic system, favouring symmetric bond lengths $R_1=R_2$ and a collinear configuration $\Theta=\pi$.\par 
We determine the vibrational states of the adiabatic PES by solving the vibrational Hamiltonian for a non-rotating triatomic molecule (see appendix~\ref{sec:3d-vib-ham}). We employ a finite difference approach~\cite{LeVegue2007Finite}, whereby the Hamiltonian is discretised on a three-dimensional grid. The block structure of the matrix Hamiltonian is such that it can be most efficiently constructed when represented on a cubic grid of edge length $N$. We find that $N=42$ is sufficient to ensure convergence of the vibrational eigenenergies to the scale of a few MHz. Reduced densities of the vibrational ground-state are shown in figure~\ref{fig:2d-pes}~(c) and (d). We find that the well exhibits bending excitations in $\Theta$ as well as symmetric stretching excitations in $R_1$ and $R_2$, whose first few excitations have spacings of 100~MHz and 135~MHz, respectively. We did not find any anti-symmetric stretching excitations among these states.\par
In addition to the approach described above, we also determined the vibrational states of the symmetric stretch Hamiltonian~\eqref{eq:vib-ham-1D}. This is an effective 1D Hamiltonian derived from the full vibrational Hamiltonian~\eqref{eq:vib_hamiltonian} for symmetric linear configurations of the nuclei and takes the form:
\begin{equation}~\label{eq:vib-ham-1D}
    H_n^{\text{eff}} = -\frac{1}{m}\frac{\partial^2}{\partial R^2} + \varepsilon_{i}(R),
\end{equation}
where $m$ is the mass of $^{87}$Rb and $\varepsilon_{i}(R)$ is a one-dimensional slice of the $i^{\text{th}}$ adiabatic PES for $R_1=R_2=R$ and $\Theta=\pi$. Using this approach, we find that the binding potential supports sixteen symmetric stretching modes. The probability density of these eigenstates are shown in figure~\ref{fig:pes}~(b). 
The states have a typical spacing of 250~MHz, which is greater than the energy spacing between states in the 3D model~\eqref{eq:vib_hamiltonian}. This is expected due to the reduced dimensionality of the binding potential in the effective model. The difference in energy of the first few symmetric stretching excitations relative to those obtained with the 3D model is 0.2\% or less (see appendix~\ref{sec:comparison}).\par
Finally, we remark that the electronic state of the binding potential in figure~\ref{fig:pes}~(b) possesses some low-$l$ character. Specifically, at the position of the local minimum the total contribution of $S$- and $D$-states is approximately 15\%. Hence, the bound states can in principle be accessed via dipole-allowed transitions (see appendix~\ref{sec:trimer-l-character}).\par
\section{Stability}\label{sec:decay}
We now consider different mechanisms which may limit the lifetime of the trimer states below the radiative lifetime of the Rydberg atom.
\begin{figure}
    \centering
    \includegraphics[width=0.475\textwidth]{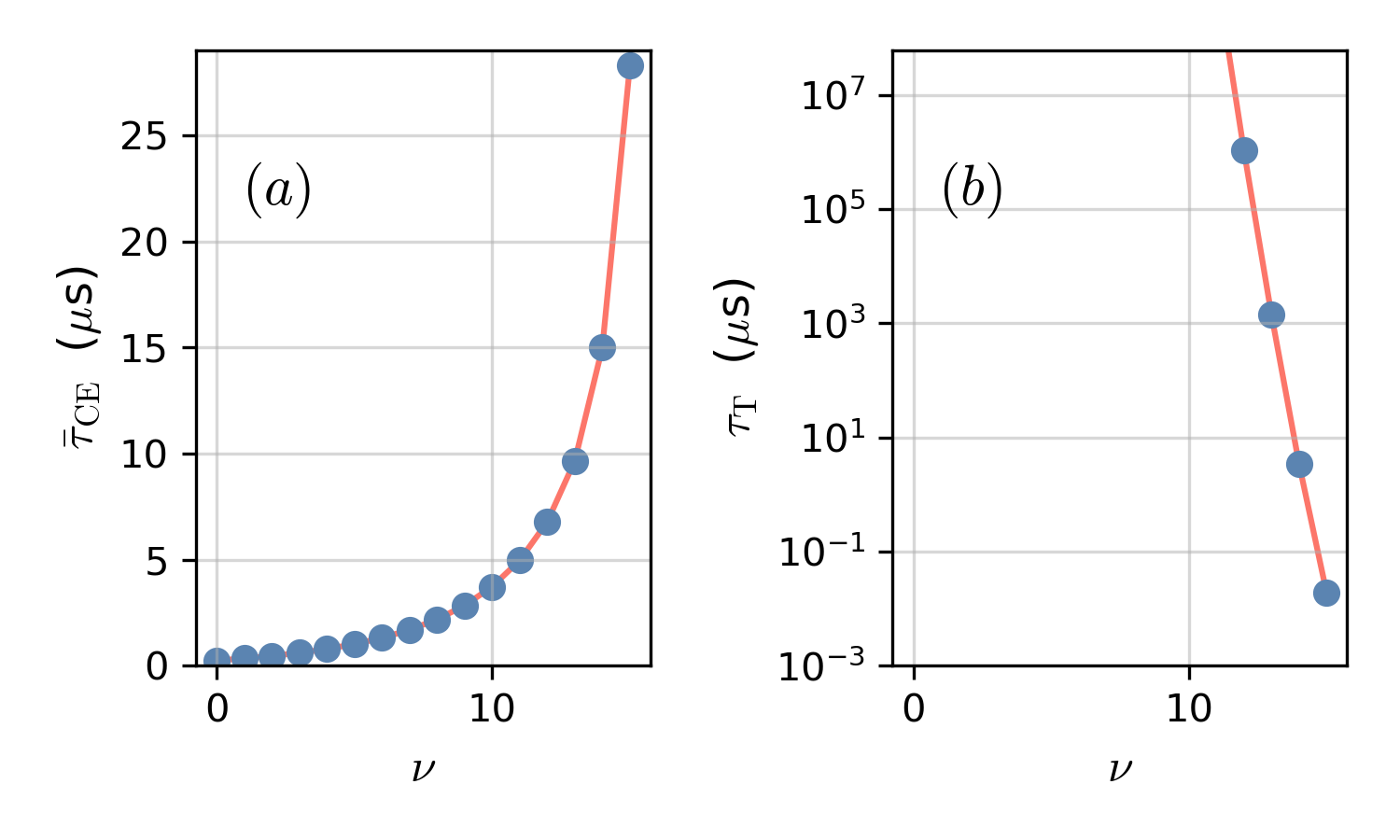}
    \caption{
    (a) Expected times for charge exchange for the symmetric stretching states in the binding potential in figure~\ref{fig:pes}~(b), with $\nu=0$ corresponding to the vibrational ground-state. The times have been averaged over the probability density distribution of the vibrational states along $R$. 
    (b) Tunnelling time of the vibrational states through the barrier in the adiabatic PES (see figure~\ref{fig:pes}~(b)).}
    \label{fig:decay-timescales}
\end{figure}
First, we examine leakage of the vibrational wavepackets from the binding potential to neighbouring PES via non-adiabatic couplings. We define the rate of non-adiabatic transitions as $\gamma_{\text{NAD}}=\Delta E \,P_{\text{LZ}}$, where $\Delta E$ is the average spacing of the vibrational states in the binding potential and $P_{\text{LZ}} = \exp(-2\pi \Gamma)$ is the Landau-Zener formula giving the probability of a non-adiabatic transition~\cite{Zener1932Nonadiabatic}. The exponent $\Gamma = \Delta^2/\alpha\dot{R}$ depends on the gap $\Delta$ and gradient $\alpha$ parameters of the diabatic curves fitted from the numerically-obtained adiabatic PES, in addition to the speed of the wavepacket $\dot{R}$ at the crossing point $R_{\text{cross}}$ of the diabatic curves. 
Taking the maximum possible value of $\dot{R}$ for a bound wavepacket, set by the energy difference between the maximum height of the binding potential's barrier and the crossing point of the diabatic curves, we find $\gamma_{\text{NAD}}\sim 10^{7}-10^{8}$~s$^{-1}$ for decay to neighbouring adiabatic curves $A$, $B$ and $C$ (see inset of figure~\ref{fig:pes}~(b)). However, if we instead define $\dot{R}$ as the speed of the highest-energy vibrational state at the crossing point $v_{\text{WKB}}=\sqrt{2[E_{15}-V(R_{\text{cross}})]/\mu}$, where $\mu$ is the reduced mass of $^{87}$Rb, the only significant decay channel is to curve $A$ with $\gamma_{\text{NAD}}\sim 10^{4}$~s$^{-1}$. We see therefore that the decay rates are highly sensitive to the wavepacket's speed. However, given the large number and complex interwoven form of the adiabatic curves, the applicability of this two-channel semi-classical treatment is unclear. In light of this, a more detailed study of the non-adiabatic coupled-channel equations with vibronic couplings \cite{Hummel2023Vibronic} would be an interesting extension of the current work.\par
We now consider the rate at which the vibrational wavepackets tunnel through the barrier shown in figure~\ref{fig:pes}~(b), which leads to a Coulomb explosion. The wavepacket tunnelling rate is defined as $\gamma_{\text{T}} = \Delta E \,  P_{\text{T}}(E)$, where $P_{\text{T}}(E)$ is the semi-classical Wentzel–Kramers–Brillouin (WKB) tunnelling probability for a wavepacket of energy $E$ (see appendix~\ref{sec:wkb}). The tunnelling times $\tau_{\text{T}}= 1/\gamma_{\text{T}}$ for the bound states are shown in figure~\ref{fig:decay-timescales}~(b). Tunnelling is only relevant for the highest-energy states, where it reaches values as small as tens of nanoseconds.\par
Transfer of charge between the ions in the system may also limit the lifetime of the trimer states. We model this process by determining the rate at which the Rydberg electron tunnels through the barrier in the Coulomb potential (see inset of figure~\ref{fig:intro-fig}~(b)). We define the charge exchange rate as $\gamma_{\text{CE}}(R) = 2 f P_{\text{CE}}(R)$ which depends on the orbital frequency of the Rydberg electron $f$ and the WKB tunnelling probability for an electron of energy $\varepsilon_{\nu}(R)$ associated with the $\nu^{\text{th}}$ PES (see appendix~\ref{sec:wkb} for further details). The factor of two accounts for the fact that the Rydberg electron can tunnel through either Coulomb barrier created by the ion pair. We show the charge exchange time $\tau_{\text{CE}}= 1/\gamma_{\text{CE}}$ as a function of $R$ in figure~\ref{fig:pes}~(b). Its value is highly sensitive to $R$, decreasing by almost three orders of magnitude from the point of the local maximum of the barrier to the minimum of the binding potential, where it is less than $1~\upmu$s. The expected charge exchange times $\bar{\tau}_{\text{CE}}$ for the bound states are shown in figure~\ref{fig:decay-timescales}~(a). For the vibrational ground-state, the expected time is approximately 0.24~$\upmu$s, which is far smaller than the radiative lifetime of the Rydberg state. The expected time increases for the excited states, since they show increasing probability density around the outer classical turning point of the binding potential (see figure~\ref{fig:pes}~(b)).\par
\begin{figure}
    \centering
    \includegraphics[width=0.475\textwidth]{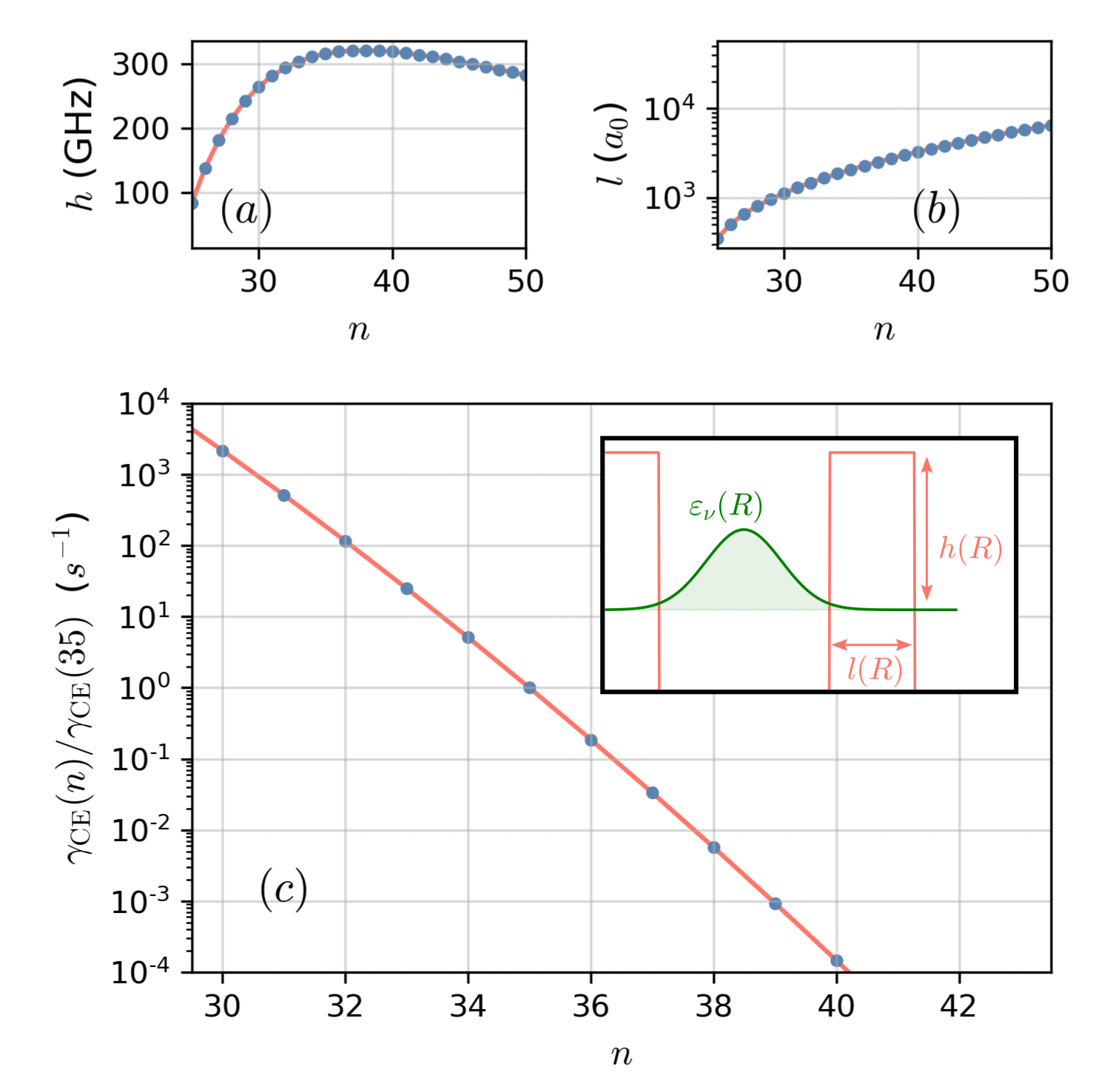}
    \caption{(a),(b) The scaling of the tunnelling barrier height $h(R_w) = V_b(R_w)-\varepsilon(R_w)$ and width $l(R_w)=z_2(R_w)-z_1(R_w)$ with $n$ (see text for definitions and inset in (c)). (c) The scaling of the charge exchange rate as a function of $n$ relative to the rate at $n=35$. The inset shows a schematic of the Rydberg electron in the Coulomb potential modelled as a box potential.}
    \label{fig:decay-rate-scaling}
\end{figure}
\begin{figure}
    \centering
    \includegraphics[width=0.475\textwidth]{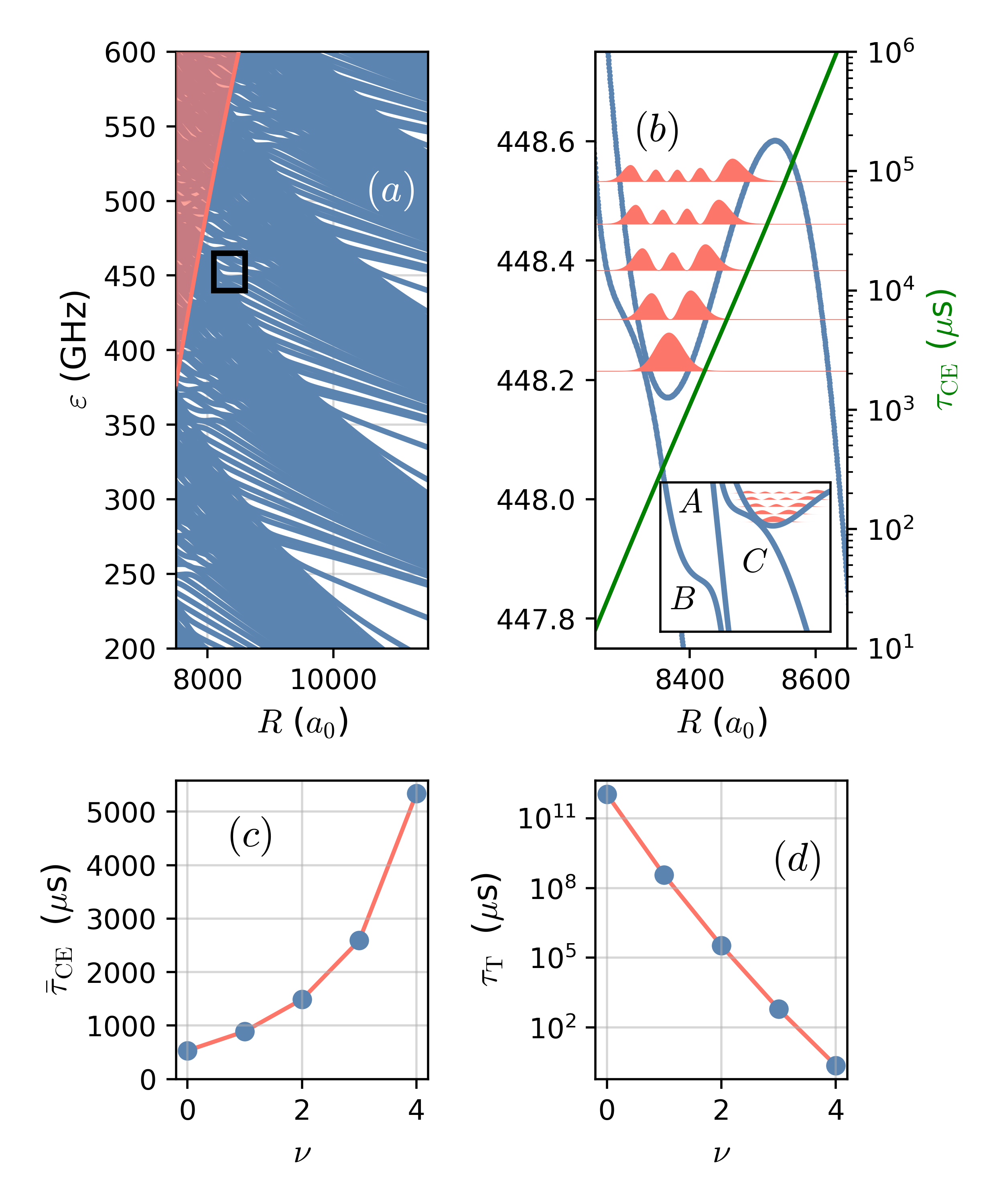}
    \caption{Adiabatic PES near the $n=38$ hydrogenic manifold. (a) 1D slice along PES for the symmetric linear configuration $R_1 = R_2 = R,\;\Theta=\pi$. The shaded region indicates the breakdown of the localised electron ansatz. (b) Magnification of the box in (a). The filled curves show the probability density of the symmetric stretching modes determined from the effective 1D vibrational Hamiltonian~\eqref{eq:vib-ham-1D}. The solid green line on the second $y$-axis is the charge exchange time of the Rydberg electron. Inset shows the three neighbouring adiabatic PES of the binding potential, labelled $A-C$. Energies given in GHz relative to the 38H atomic Rydberg state. (c) Expected times for charge exchange for the symmetric stretching modes in (b). 
    (d) Tunnelling time of the vibrational symmetric stretch modes through the barrier visible in (b).}
    \label{fig:n38-results}
\end{figure}
Concluding our discussion of the well at $n=35$, we have found that tunnelling of the Rydberg electron will limit the lifetime of the low-energy bound states, whilst the lifetime of the highest-energy states is limited by wavepacket tunnelling. Given these processes, we expect lifetimes on timescales of 0.1-10~$\upmu$s for the trimer states at $n=35$.\par
How might these expected lifetimes change at higher $n$? We anticipate that rates of vibrational wavepacket tunnelling will increase, since the binding potentials become shallower at larger $n$. To investigate how the rate of charge exchange varies with $n$, we model the Coulomb barrier experienced by the Rydberg electron as a box potential of height $h(R) = V_b(R) - \varepsilon_A^{(1)}(R)$ and width $l(R) = z_2(R)-z_1(R)$ (see inset of figure~\ref{fig:decay-rate-scaling}~(c)). $z_2$ and $z_1$ are classical turning points of the Rydberg electron at position $z$ in the Coulomb potential $V(z,R)$ defined at internuclear separation $R$. The WKB tunnelling rate~\eqref{eq:wkb-wavepacket} of the Rydberg electron is then given by $\gamma_{\text{CE}} = 2f\exp[-2\sqrt{2h(R)}l(R)]$. The Rydberg orbital frequency $f$ scales proportional to $n^{-3}$~\cite{Gallagher2003Back}.\par 
Taking $R=R_w$, the $n$-scaling of $h(R_w)$, $l(R_w)$ and $\gamma_{\text{CE}}$ can be determined from equations~\eqref{eq:energy-correction} and~\eqref{eq:Rw}, which we show in figure~\ref{fig:decay-rate-scaling}.
We see that $\gamma_{\text{CE}}$ is monotonously decreasing and highly sensitive to the value of $n$, ranging over nearly eight orders of magnitude between $n=30$ and $n=40$. In particular, we find that the decay rate at $n=38$ to be just over two orders of magnitude smaller than at $n=35$. We thus expect charge exchange for vibrational states at $n=38$ to occur on timescales of 10-100~$\upmu$s, which are more comparable with the radiative lifetime of the Rydberg atom.\par
To confirm this, we determine adiabatic PES in the vicinity of the $n=38$ Rydberg manifold and the associated 1D symmetric stretching modes satisfying the vibrational Hamiltonian~\eqref{eq:vib-ham-1D}. We show the results in figure~\ref{fig:n38-results}. The binding potential at $n=38$ is shallower than at $n=35$ and hence supports fewer vibrationally-bound states. Despite this, tunnelling of the vibrational wavepacket remains relevant only for the highest energy states (see figure~\ref{fig:n38-results}~(d)). The expected charge transfer times $\bar{\tau}_{\text{CE}}$ shown in figure~\ref{fig:n38-results}~(c) are considerably larger than for $n=35$. For the vibrational ground-state, we predict a time of approximately 500~$\upmu$s, which is more than three orders of magnitude larger than the ground-state of the $n=35$ binding potential (cf. figure~\ref{fig:decay-timescales}~(a)). The non-adiabatic decay rates obtained with the maximum wavepacket speed are similar to those at $n=35$, whilst those determined with $v_{\text{WKB}}$ are vanishingly small for all decay channels.\par
In summary, we expect bound states of $^{87}\text{Rb}_3^{2+}$ to become longer-lived at higher $n$ due to the decreasing rate of charge transfer. However, with increasing $n$ the depth of the binding potential decreases such that eventually tunnelling of the vibrational wavepackets will become the limiting factor in the stability of the trimer.
\section{Summary \& Outlook}~\label{sec:outlook}
We predict the existence of metastable doubly-charged molecules formed due to long-range bonding between a single Rydberg atom and two cations. Although such a system is not expected to be stable in the electronic ground-state, we find that above a critical value of $n$ the Rydberg atom acquires a sufficiently large quadrupole moment to counterbalance the Coulomb repulsion between the ion pair, forming GHz-deep binding potentials bearing several vibrationally-bound states.\par
We began by considering the competing interactions within the effectively one-electron $^{87}$Rb$_3^{2+}$ system. Whilst the Rydberg electron delocalises below a critical internuclear separation $R_d$, for suitably large $n$ the ion-Rydberg interaction is strong enough to mix states in neighbouring hydrogenic manifolds at separations $R > R_d$. We found that level-repulsion between these states gives rise to a host of potential wells deep enough to support three-body bound states in the ultracold regime. Using a semi-classical approach, we assessed the stability of these bound states against non-adiabatic decay and tunnelling of the vibrationally-bound wavepackets and Rydberg electron.\par
In particular, we found that the tunnelling rate of the Rydberg electron is highly sensitive to $n$, differing by over three orders of magnitude between $n=35$ and $n=38$. At low $n$, the molecular lifetime will be limited chiefly by tunnelling of the Rydberg electron. In contrast, tunnelling of the vibrational wavepacket will limit the lifetime at high $n$ since the binding potentials become increasingly shallow. For the case of $n=38$ however, we found that the system is sufficiently stable against both tunnelling mechanisms that the molecule's lifetime should be limited only by radiative decay. We note that the responsiveness of the Rydberg electron's tunnelling rate to $n$ may make similar systems interesting for studying controlled charge transport~\cite{Mukherjee2019Charge}.\par
In general, we expect non-adiabatic effects will be present in systems such as this due to the high density of electronic states. Regimes with strong vibronic couplings may exhibit conical intersections~\cite{Hummel2021Synthetic} or spontaneous symmetry-breaking~\cite{Becker2024Synthetic}, the latter of which is unique to polyatomic systems~\cite{Bersuker2021Jahn}. Investigations of such regimes are left to future work.\par
Although the molecular bound states are accessible with dipole-allowed transitions, bringing the three atoms together in order to photoassociate the molecule poses a unique experimental challenge. Optical tweezer setups may be advantageous for this purpose, since they offer precise control over interparticle separation and can be paired with single-atom laser addressing schemes. Beginning with three trapped neutral atoms in their ground-state, one could create a pair of cold free-floating ions using schemes demonstrated in recent experiments~\cite{Dieterle2021Transport,Zuber2022Observation,Berngruber2024Insitu} and then immediately drive a Rydberg transition of the remaining atom. The successful formation of the trimer state could be confirmed by an absent or delayed Coulomb explosion. Alternatively, the binding could be verified through mass spectroscopic measurements, similar to the approach already employed for ion-Rydberg dimers~\cite{Zuber2022Observation}.\par
Despite the practical challenges in observing these unusual molecules, their existence within the parameter space of an otherwise familiar system underscores once more the rich physics which can emerge from exploring the internal structure of atoms.\\
\section*{Acknowledgements}
The authors express their gratitude to Frederic Hummel for discussions during the formative stages of the project. D.J.B. also thanks Rick Mukherjee for his insights related to charge transport. This work is funded by the Cluster of Excellence “Advanced Imaging of Matter” of the Deutsche Forschungsgemeinschaft (DFG)-EXC 2056, Project ID No. 390715994.
%
\appendix
\section*{Appendices}
\section{Triatomic vibrational Hamiltonian}\label{sec:3d-vib-ham}
For a non-rotating triatomic molecule, the vibrational Hamiltonian is given by~\cite{Carter1982Variational,Handy1987Derivation}:
\begin{widetext}
\begin{equation}\label{eq:vib_hamiltonian}
    \begin{split}
        H_n &= \frac{1}{m} \bigg[ -\frac{\partial^2}{\partial R_1^2} -\frac{\partial^2}{\partial R_2^2} -\cos(\Theta)\frac{\partial}{\partial R_1}\frac{\partial}{\partial R_2} \bigg]\\
        &-\frac{1}{m}\bigg(\frac{1}{R_1^2}+\frac{1}{R_2^2}-\frac{\cos(\Theta)}{R_1 R_2} \bigg)\bigg( \frac{\partial^2}{\partial\Theta^2} +\cot(\Theta)\frac{\partial}{\partial\Theta}\bigg)\\
        &-\frac{1}{m}\bigg( \frac{1}{R_1 R_2}-\frac{1}{R_2}\frac{\partial}{\partial R_1} - \frac{1}{R_1}\frac{\partial}{\partial R_2}\bigg)\bigg(\cos(\Theta) + \sin(\Theta)\frac{\partial}{\partial\Theta} \bigg)\\
        &+\varepsilon_{\nu}(R_1,R_2,\Theta),\\
    \end{split}
\end{equation}
\end{widetext}
where $m$ is the atomic mass of $^{87}$Rb and $\varepsilon_{\nu}(R_1,R_2,\Theta)$ is the $\nu^{\text{th}}$ adiabatic PES. This Hamiltonian acts on wavefunctions $\chi(R_1,R_2,\Theta)$ normalised as $\int d R_1 d R_2 d \Theta \sin\Theta |\chi(R_1,R_2,\Theta)|^2=1$.\par
\section{Comparing vibrational eigenenergies}\label{sec:comparison}
Table~\ref{tab:energy_comp} compares the energies of the first five symmetric stretching excitations obtained with the full 3D vibrational Hamiltonian~\eqref{eq:vib_hamiltonian} and the 1D effective model~\eqref{eq:vib-ham-1D}. We find a relative energy difference of 0.2\% or less between the eigenenergies.
\begin{table}[t]
    \centering
    \begin{tabular}{c|c|c|c}
        \hline
        $\nu$ & 3D model~\eqref{eq:vib_hamiltonian} & 1D model~\eqref{eq:vib-ham-1D} & \% rel. diff.\\\hline
        0 & 414.67 & 413.81 & 0.21 \\\hline
        1 & 414.81 & 414.12 & 0.17 \\\hline
        2 & 414.94 & 414.41 & 0.13 \\\hline
        3 & 415.07 & 414.68 & 0.09 \\\hline
        4 & 415.20 & 414.68 & 0.06 \\\hline
    \end{tabular}
    \caption{Comparison of energies of the first five symmetric stretching eigenstates (including the ground-state) of the $^{87}$Rb$_3^{2+}$ system obtained with the full (3D) and effective (1D) models. Values are given to two decimal places.}
    \label{tab:energy_comp}
\end{table}
\section{Semi-classical tunnelling model}\label{sec:wkb}
The probability for a wavepacket of mass $m$ with energy $\epsilon$ to tunnel through a 1D barrier $V(z)$ is given in the Wentzel-Kramers-Brillouin (WKB) approach~\cite{Friedrich2017Theoretical,Griffiths2018Introduction} by:
\begin{equation}~\label{eq:wkb-wavepacket}
    P = \exp\bigg[  -2\sqrt{2m} \int_{z_1}^{z_2} dz \sqrt{V(z)-\epsilon}  \bigg].
\end{equation}
The limits of the integral $z_1$ and $z_2$ are the inner and outer classical turning points of the barrier, such that $\epsilon = V(z_1) = V(z_2)$.\par
For the case of wavepacket tunnelling, $m$ is given by the reduced mass of $^{87}$Rb and $\epsilon$ corresponds to the energy of the vibrationally-bound state determined from the symmetric stretching Hamiltonian~\eqref{eq:vib-ham-1D}. For the case of electron tunnelling, $m$ is the Rydberg electron's mass $m_e$ and $\epsilon$ is defined by the $R$-dependent adiabatic PES shown in figures~\ref{fig:pes}~(b) and~\ref{fig:n38-results}~(b). The charge exchange rate $\gamma_{\text{CE}}$ is then determined by multiplying the result of~\eqref{eq:wkb-wavepacket} with the Kepler frequency of the Rydberg electron's orbit $f = 1/T$, where $T$ is the orbital period. For $n=35$, we find $f\sim10^9$~Hz.
\section{Photoassociation of $^{87}\text{Rb}_3^{2+}$}\label{sec:trimer-l-character}
\begin{figure}[H]
    \centering
    \includegraphics[width=0.475\textwidth]{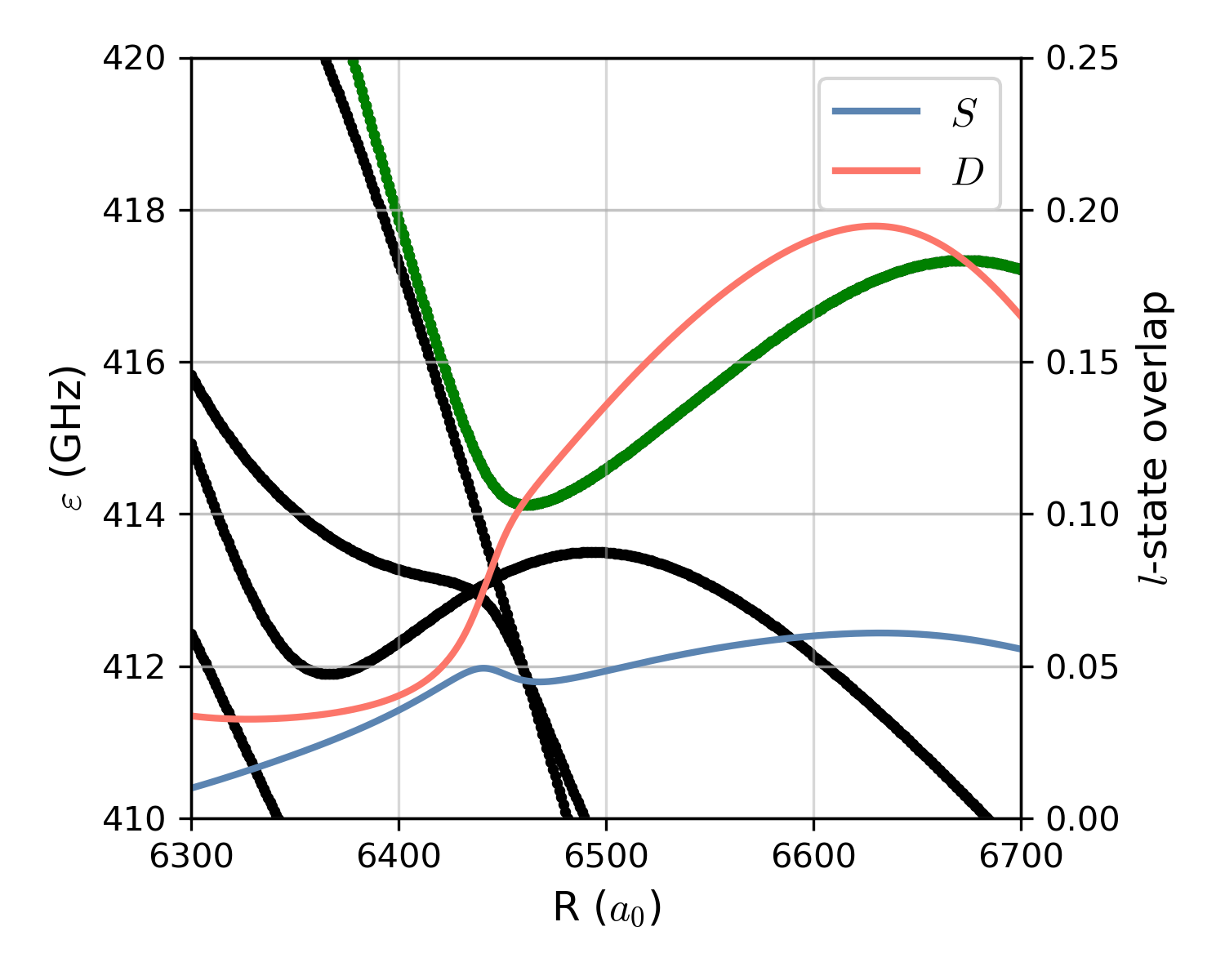}
    \caption{Adiabatic PES near the $n=35$ hydrogenic manifold. The secondary axis shows the overlap of the electronic state of the binding potential (green) with $S$- and $D$-state atomic Rydberg states.}
    \label{fig:l-char}
\end{figure}

We see from figure~\ref{fig:l-char} that the electronic state of the $^{87}\text{Rb}_3^{2+}$ binding potential exhibits significant overlap with low angular momentum states ($S$- and $D$-states) such that photoassociation within the framework of single- or two-photon transitions should be possible to produce $^{87}\text{Rb}_3^{2+}$ states.\\
\indent For the symmetric linear configuration considered in figure~\ref{fig:l-char}, there is a vanishing contribution from $P$-states due to the fact that the leading-order term is a charge-quadrupole interaction. Due to Clebsch-Gordon selection rules, Rydberg states with angular momentum differing only by one are not coupled by such interactions such that the parity of the angular momentum character of the electronic states is preserved. Away from this highly-symmetric case however, parity symmetry is broken.
\end{document}